\documentclass[12pt,twocolumn]{article}
\usepackage[a4paper,left=20mm,right=20mm,top=25mm,bottom=25mm,includeheadfoot]{geometry}

\usepackage{mathpazo}
\usepackage{graphicx}
\usepackage{amsmath,textcomp}
\usepackage{makeidx}
\makeindex

\usepackage{sectsty}
	\allsectionsfont{\sffamily\raggedright}
	\sectionfont{\sffamily\large\raggedright}
	\subsectionfont{\sffamily\normalsize\raggedright}
\usepackage{ftnright}

\usepackage{widetext}
\usepackage{flushend}
\usepackage{cuted}
\usepackage{color}
\usepackage{graphicx}
\usepackage{hyperref}
\usepackage{pstricks,amssymb}

\usepackage{fancyhdr}
\begin{document}
\title{\vspace{-2em}\bfseries\sffamily A heuristic derivation of radiative power loss and radiation reaction from the kinetic power of electric inertial mass of a charge}
\author{\normalsize Ashok K. Singal\\[2ex]
Astronomy and Astrophysics Division, Physical Research Laboratory\\
Navrangpura, Ahmedabad 380 009, India.\\
{\tt ashokkumar.singal@gmail.com}
}
\date{\itshape Submitted on 08-OCT-2016}
\maketitle

\thispagestyle{fancy}

\begin{abstract}
{\sffamily
It is shown that formulas for the radiative power loss and radiation reaction from a charge can be derived in a heuristic manner from the kinetic power (rate of change of the kinetic energy) of its electric inertial mass. 
The derivation assumes a non-relativistic but otherwise an arbitrary motion of the charge. We exploit the fact that as the charge 
velocity changes because of a constant acceleration, there are accompanying modifications in its electromagnetic fields 
which can remain concurrent with the charge motion because the velocity as well as acceleration information enters into the field expression.  
However, if the acceleration of the charge is varying, information about that being not present in the field expressions, 
the electromagnetic fields get ``out of step'' with the actual charge motion. Accordingly we arrive at a radiation reaction formula for an 
arbitrarily moving charge, obtained hitherto in literature from the self-force, derived in a rather cumbersome way from the 
detailed mutual interaction between various constituents of a small charged sphere. This way we demonstrate that a power 
loss from a charge occurs only when there is a change in its acceleration and the derived instantaneous power loss is directly proportional 
to the scalar product of the velocity and the rate of change of the acceleration of the charge. 
}\\ 
\hrule
\end{abstract}

\section{Introduction}
The radiation reaction was first proposed by Lorentz~\cite{lor92, lor04} and later Abraham \cite{abr05} and Lorentz~\cite{16} derived it in detail 
for an arbitrarily moving small charged sphere, and is available in various forms in many text-books 
\cite{1,2,3,25,7,24,20}. The formula, hitherto obtained in literature in a rather cumbersome way, is evaluated from the 
detailed mutual interaction between constituents of the charged sphere. The self-force turns out to be proportional to the rate 
of change of acceleration, independent of the radius of the small sphere. 
One obtains the instantaneous radiative power loss formula by a scalar product of the self-force with the velocity of the charge. 
The same formula for the radiative power loss is also obtained from the Poynting flux in the neighbourhood of a ``point charge'' 
in arbitrary motion \cite {p32a}. Further, the radiation reaction formula has been derived also from the rate of electromagnetic momentum 
flow, calculated using the Maxwell stress tensor, across a surface surrounding the  neighbourhood of a point charge, \cite {p32}.

However this power loss formula does not agree with the standard Larmor's formula,  
where one calculates the Poynting flux through a spherical surface of large enough radius $r$ centred on the time-retarded position of the charge. The flux turns out to be proportional 
to the square of  acceleration ($\propto \dot{{\bf v}}^2$) \cite{1,2,25}. 

There is extensive literature on this controversy of which of these two 
formulas gives correct description of radiative losses \cite{6,50,51,56}. 
Larmor's formula leads to wrong conclusions in the instantaneous rest-frame of an accelerated charge,  
where the charge has no velocity and thus no kinetic energy to be lost into radiation. On the other hand Larmor's formula predicts a 
continuous radiative loss proportional to the square of acceleration even for an instantly stationary charge.
Often an extraneous acceleration-dependent term called Schott energy is introduced to make the two formulas conform to each other \cite{7,row10,57,44,41,steane}.
But recently it has been explicitly shown that the Poynting flux passing through a spherical surface of vanishingly  
small dimensions surrounding the charge, in its instantaneous rest-frame, is zero \cite {p32a}. 
Actually in all neighbourhood of the charge in its instantaneous rest-frame, the transverse terms of the time-retarded velocity fields 
cancel the acceleration fields, which were responsible in Larmor's formula for radiation.  
This removes the need for the acceleration-dependent Schott-energy term, introduced in the literature on an ad hoc 
basis to comply with law of energy conservation (see also  \cite {p32b,p32c,p58b}).

The radiation reaction formula could also be derived from Larmor's formula of radiative losses using the law of energy conservation 
\cite{1,2,25}, but only if the Schott term remains unchanged at the ends of the time interval considered, which happens if the motion is cyclic. 
Momentum conservation also remains a problem in Larmor's radiation picture. The radiation pattern of an accelerated charge has a $\sin^2\phi$ dependence 
about the direction of acceleration \cite{1,2,25}. Due to this azimuthal symmetry the net momentum carried by the 
radiation is nil. Therefore the charge too cannot be losing momentum, even though it is undergoing radiative losses. 
Thus we have a paradox of a radiating charge losing its kinetic energy but without a corresponding change in its linear momentum. 
An effect of this inconsistency appears in a synchrotron radiation case, where Larmor's formula leads to conclusions about the dynamics of the 
radiating charge that are not conversant with special relativity and violate energy-momentum conservation\cite {p58a}. However no such inconsistencies arises when one makes use of the radiation reaction formula to calculate energy-momentum losses of the radiating charge \cite {p30}.

More recently it has been shown that the two formulas are compatible and no controversy really arises if one keeps a 
proper distinction between the retarded time and the real time \cite{p52}. In particular, one gets Larmor's formula, 
with radiative losses proportional to the square of the acceleration if one expresses the radiated power
in terms of quantities describing motion of the charge at the retarded time. On the other hand if the motion of the charge is expressed 
in terms of real time (``present'') quantities, then one arrives at the power loss formula usually derived from the radiation reaction formulation, 
i.e., the radiative power loss proportional to the scalar product of the velocity and the first time derivative of the acceleration 
of the charge. 

Without going any further into the controversy between the two radiation loss formulas, 
here we show that the radiation reaction formula can be derived in an alternate, though heuristic, method, from the mechanical motion of the charge if one takes the 
electrical mass of the charge as its inertial mass. For this we shall make use of the kinetic power, i.e., a temporal rate of change of the kinetic 
energy, of the charge to derive a formula for radiative losses of the charge.
\section{Electric inertial mass of a charge}
The electromagnetic field momentum is given by the volume integral \cite{1} 
\begin{equation}
\label{eq:1f1b}
{\bf P_{\rm field}}=\frac{1}{4\pi c}\int{{\rm d}V}\:({\bf E}\times{\bf B}).
\end{equation}
A charge, assumed to be a uniformly charged spherical shell of radius 
$r_{\rm o}$, moving with a non-relativistic, uniform velocity $\bf v_{\rm o}$, from the above volume integral, possesses an electromagnetic field momentum 
\begin{eqnarray}
\label{eq:1f1a}
{\bf P_{\rm field}}=\frac{2e^2{\bf v}_{\rm o}}{3r_{\rm o}\: c^{2}}=m_{el} {\bf v_{\rm o}}\:,
\end{eqnarray}
with an electric mass defined as $m_{el}=4U_{0}/3c^{2}$ \cite{29}, 
where $U_{0}=e^2/2r_{\rm o}$ is the energy in self-fields of the charge in its rest frame. The factor of $4/3$ in the inertia of electric mass has long since been highly annoysome. Poincar\'{e} \cite{34} pointed out that 
in a real charged particle, there must be some non-electrical (!) ``binding'' forces  to balance the Coulomb self-repulsion 
of the charge, which would remove the factor of 4/3. However these {\em non-electrical} binding forces are not represented   
in the expressions of the {\em electromagnetic} fields and an explanation for this factor of 4/3 must be found within the electromagnetic 
theory itself. It has been explicitly shown \cite{p1} 
that this extra factor in the expression for the total electromagnetic momentum of the charge arises 
because of the energy flow associated with the electromagnetic self-repulsion force within the charge constituents. 
The net force on one hemisphere of the charge is along the direction of motion, and on the remaining hemisphere it is  
in a direction opposite to the motion. Therefore as the charge moves, a positive work is being done by self-force {\em on} the forward hemisphere, 
while an equal amount of work is being done {\em by} the backward hemisphere against the self-force. 
Though there is no net increase in the energy of the total system, yet because of the electromagnetic self-force there is a continuous flow of energy 
across the charged sphere between its two halves, implying a corresponding momentum due to this energy-flow. This momentum is 
important even for non-relativistic velocities and gives 1/3rd additional contribution to the otherwise momentum of 
the charge \cite{p1}, thereby explaining this intriguing factor of 4/3 in the total electromagnetic momentum. 
\section{Electromagnetic field momentum of a uniformly accelerated charge}
Electromagnetic field of a charge $e$, from the laws of
electrodynamics, is determined at time $t$ by the charge motion (${\bf v}$ and $\dot{\bf v}$) at the retarded time $t'=t-r/c$ \cite{1,2,25,p45}. 
\begin{eqnarray}
\nonumber
{\bf E}&=&\left[\frac{e({\bf n}-{\bf v}/c)} {\gamma ^{2}r^{2}(1-{\bf n}\cdot{\bf v}/c)^{3}}\right.\\
\label{eq:1}
&& + \left.\frac{e{\bf n}\times\{({\bf n}-{\bf v}/c)\times \dot{\bf v}\}}{rc^2\:(1-{\bf n}\cdot {\bf v}/c)^{3}}\right]_{t'}\\
{\bf B}&=& {\bf n}\times {\bf E}
\end{eqnarray}
The first term on the right hand side of (Eq. (\ref{eq:1})) that fall with distance as $1/r^2$, is called velocity fields while the second term, falling with distance as 
$1/r$, is called the acceleration fields, the latter generally assumed to be solely responsible for radiation from the charge.
It is a standard practice to assign the Poynting flux, calculated using acceleration fields, through a spherical surface, say $\Sigma$, of radius 
$r=c(t-t')$, centred on the charge position at the retarded time $t'$, as the radiation losses by the charge at time $t'$, to get 
Larmor's formula for radiative losses \cite{1,2,25}. However, Poynting theorem tells us that the rate of the kinetic energy loss by charge at 
{\em present time} $t=t'+r/c$ (and not at retarded time $t'$) is related to the instantaneous outgoing electromagnetic power (Poynting 
flux) at $t$ from the surface $\Sigma$ \cite{1,2,25}. It may though be recalled that the fields at the surface $\Sigma$ are determined 
by the motion of the charge at the retarded time $t'$ (Eq.~(\ref{eq:1})). This might appear to be a break down of causality, after all,
how come the Poynting flux  determined from the motion of the charge in past, i.e., at an earlier time $t'$, is being equated to the 
kinetic energy-loss rate of the charge at a later time $t$? How can one be sure that the charge will not behave
erratic between $t'$ and $t$, thus while keeping the Poynting flux unaffected (which is already decided by the charge motion at $t'$)
but modifying the kinetic power loss rate of the charge? Actually, even the charge motion at the present time $t$ follows from the charge 
motion at $t'$, determined by the laws of mechanics, and thus both the electromagnetic fields on the surface $\Sigma$ as well as the 
charge motion at time $t$ are determined by the charge motion at $t'$ and there is no conflict with the causality.

Let us first consider the case of a uniform acceleration $\dot{\bf v}$. The charge motion due to acceleration is,  
\begin{eqnarray}
\label{eq:41b}
{\bf v}_{\rm o}&=&{\bf v}+ \dot{{\bf v}} (t_{\rm o}-t')\\
\label{eq:41c}
\dot{{\bf v}}_{\rm o}&=&\dot{{\bf v}}\:,
\end{eqnarray}
were ${\bf v}$, $\dot{\bf v}$ on the right hand side 
represent respectively the velocity and acceleration 
of the charge at the retarded time $t'$, while ${\bf v}_{\rm o}, \dot{{\bf v}}_{\rm o}$ on the left hand side 
represents the corresponding values at the present time $t$. 

For our considerations, we assume the charge motion to be non-relativistic, where Lorentz contraction may not play any role and the 
moving charge continues to be a uniformly charged sphere of radius $r_{\rm o}$, as when it is at rest. 

Using the vector identity ${\bf v}={\bf n}({\bf v}.{\bf n}) - {\bf n}\times\{{\bf n}\times{\bf v}\}$ in the expression for electric field (Eq. (\ref{eq:1})), the electric field for a non-relativistic motion of the charge, thereby dropping all terms which are non-linear in $\bf v$ or its derivatives, can be written as 
\begin{eqnarray}
\nonumber
{\bf E}&=&\frac{e\:{\bf n}}{r^2(1-{\bf n}\cdot{\bf v}/c)^2}+\frac{e{\bf n}\times({\bf n}\times {\bf v})} {cr^{2}} \\
\label{eq:1a}
&&+ \frac{e{\bf n}\times({\bf n}\times \dot{\bf v})}{c^2r}\:.
\end{eqnarray}

Now, in the case of a uniform 
acceleration, the {\em retarded value} of the velocity will be ${\bf v} = {\bf v}_{\rm o} -\dot{\bf v} r/c$ (Eq. (\ref{eq:41b})). Then Eq.~(\ref{eq:1a}) for the electric field becomes 
\begin{eqnarray}
\nonumber
{\bf E}&=&\frac{e{\bf n}}{r^{2}}\bigg[1
+\frac{2{\bf n} \cdot {\bf v}_{\rm o}}{c}-\frac{2{\bf n} \cdot \dot{\bf v}r}{ c^2}\bigg]\\
\label{eq:1a11}
&&+\frac{e{\bf n}\times({\bf n}\times {\bf v}_{\rm o})} {cr^{2}}\:,
\end{eqnarray}
with the magnetic field given by
\begin{equation}
\label{eq:12k}
{\bf B}= \frac{-e\:{\bf n}\times {\bf v}_{\rm o}}{r^{2}c}\:.
\end{equation}
This begets for the Poynting flux  
\begin{eqnarray}
\nonumber
{\cal S}&= &\frac{c}{4\pi}\oint_{\Sigma}{{\rm d}\Sigma}\:{\bf n} \cdot ({\bf E}\times {\bf B})\\
\nonumber
&=&\frac{e^2{\bf v}_{\rm o}^2}{2 r^2c}\int_{\rm o}^{\pi} {\rm d}\theta\: \sin^3\theta\\
\label{eq:11c1}
&=& \frac{2e^{2}{\bf v}_{\rm o}^{2}}{3r^2c}\:.
\end{eqnarray}
In the case of a uniformly accelerated charge, evidently, there is no term proportional to $\dot{\bf v}^{2}$, 
independent of $r$, which is usually called the radiated power. Instead, the Poynting flux (Eq.~(\ref{eq:11c1})) is merely what would be for a hypothetical charge moving with a uniform velocity ${{\bf v}}_{\rm o}$, which is nothing but the velocity of the  actual charge at the present time. 

Now in the instantaneous rest-frame of the charge, ${\bf v}_{\rm o}=0$, which means Poynting flux is zero (Eq.~(\ref{eq:11c1})). In fact, everywhere, the transverse component of the electric field is zero, 
and so is the magnetic field. Incidentally Pauli \cite{33} first pointed it out that magnetic field 
is throughout zero in the instantaneous rest-frame of a uniformly accelerated charge, indicating the absence of radiation 
from a uniformly accelerated charge.

We can substitute for ${\bf E}$ and ${\bf B}$ from Eqs.~(\ref{eq:1a11}) and (\ref{eq:12k}) in Eq.~(\ref{eq:1f1b}) to calculate the electromagnetic field momentum of a uniformly accelerated charge, having a non-relativistic motion. The transverse component of the electric field makes a nil contribution to the volume integral in Eq.~(\ref{eq:1f1b}). 
In fact, the only finite contribution to the electromagnetic field momentum comes from the first radial term ($e{\bf n}/r^2$) in Eq.~(\ref{eq:1a11}) to yield 
\begin{eqnarray}
\nonumber
{\bf P_{\rm field}}&=&\frac{-e^2}{4\pi c}\int{{\rm d}V}\;\frac{{\bf n}\times({\bf n}\times {\bf v}_{\rm o})}{r^4c}\\
\nonumber
&=&\frac{e^2{\bf v}_{\rm o}}{2c^2}\int_{\rm o}^{\pi} {\rm d}\theta\:\sin^3\theta\int_{r_{\rm o}}^{\infty}\frac{{\rm d}r}{r^2}\\
\label{eq:1f2}
&=&\frac{2e^2{\bf v}_{\rm o}}{3r_{\rm o}\: c^{2}}\:.
\end{eqnarray}
This is the electromagnetic field momentum in the volume outside the sphere of radius $r_{\rm o}$. One gets exactly the same expression (Eq.~(\ref{eq:1f1a})) for the electromagnetic field momentum for a charge moving with a uniform velocity equal to the ``present velocity'', ${\bf v}_{\rm o}$, of the uniformly accelerated charge.  

It has been shown explicitly elsewhere that 
the self-field energy-momentum of a charge moving with a uniform velocity can be represented by the 
kinetic energy-momentum of the charge, provided its electric mass is taken as its inertial mass \cite{29,p1}. 
It has also been shown that for a uniform acceleration, the contribution of the acceleration fields to the total 
field energy of the charge is just sufficient to match exactly the amount needed for its velocity-dependent self-field energy 
based on its extrapolated motion at a future time \cite{p7}. This is possible since both make use of the velocity and acceleration of the 
charge at $t'$, and things in mechanics and electrodynamics are such that
the rates of change of energy from both at any later time $t(>t')$ remain synchronized for a uniformly accelerated charge. We have also presently shown that for a uniformly accelerated charge, but with a non-relativistic motion (see \cite{p7} for a full relativistic treatment), total 
Poynting flux, including from both velocity and acceleration field terms, at any 
time is just equal to that of a charge moving uniformly with a velocity equal to the 
instantaneous ``present'' velocity of the accelerated charge.
Further, it was shown that there is no excess flux in fields that could be treated as radiation, over and above
that implied from the instantaneous ``present'' velocity of a uniformly accelerated charge. 

It follows that in the case of a {\em uniformly} accelerated charge, its rate of change in kinetic energy is concurrent with the rate of 
change in its electromagnetic field energy ${\cal P}_{\rm field}$, and is therefore given by the scalar product of the 
rate of change of its electromagnetic field momentum $\dot{\bf P}_{\rm field}$, with its instantaneous velocity ${\bf v}_{\rm o}$.
\begin{equation}
\label{eq:1f3}
{\cal P}_{\rm field}=\dot{\bf P}_{\rm field}\cdot {\bf v}_{\rm o}=\frac{2e^2\dot{\bf v}\cdot {\bf v}_{\rm o}}{3r_{\rm o}\: c^{2}}\:.
\end{equation}

\section{Radiative losses from a charge moving arbitrarily}
As long as the charge continues to move with acceleration {\em equal to that at
the retarded time} (i.e., a uniform acceleration) no mismatch in field energy takes place. However, 
a mismatch in the field energy with respect to the kinetic energy of the charge could occur when charge moves
with a non-uniform acceleration since there is no information in the
field expressions about the rate of change of acceleration of the charge (cf. Eq. (\ref{eq:1})).
In that case the ``real'' velocity of the charge differs from the extrapolated value obtained from 
the value of acceleration at the retarded time and the kinetic energy due to 
the actual velocity no longer agrees with that determined by the acceleration at the retarded time.
Then the total energy in electromagnetic fields does not correspond to that expected in self-field because of 
the ``real'' velocity of the charge, and it is this difference in
the field energy that could be said to be the power loss due to radiation. 
Thus, in the case of a non-uniform acceleration there will be a mismatch in the field energy 
with respect to the kinetic energy, calculated from the actual velocity of the charge, since the rate of change of acceleration $\ddot{{\bf v}}$ 
does not enter in the electromagnetic field expression (Eq. (\ref{eq:1})), while it does determine the actual velocity of the charge 
(after all that is how $\ddot{{\bf v}}$ gets defined).

We consider a non-relativistic motion of a uniformly charged spherical shell of radius 
$r_{\rm o}$, moving initially with a uniform acceleration $\dot{{\bf v}}$ up to 
some time $t'$ and then a rate of change of acceleration, $\ddot{{\bf v}}$, is imposed on the charge motion. 
Now at a time $t_{\rm o}=t'+r_{\rm o}/c$ the information about the change in acceleration has not yet gone beyond 
$r_{\rm o}$, hence the electromagnetic fields and the energy-momentum in them outside the charge radius $r_{\rm o}$ are  
unaffected by the imposition of $\ddot{{\bf v}}$ on the charge motion at $t'$. Therefore the electromagnetic energy-momentum in fields 
external to the charge continues at $t_{\rm o}$ to be that of a uniformly accelerated charge, and thus determined from 
${{\bf v}}$ and $\dot{{\bf v}}$ at $t'$. Thus energy in the fields mimics the extrapolated value of the kinetic energy of 
the charge, with electric mass of the charge taken as its inertial mass, for its erstwhile uniform acceleration \cite{29,p1,p7}. However, 
due to a change in the acceleration ($\ddot{{\bf v}}$), the actual kinetic energy of the charge at $t_{\rm o}$ is no longer that determined 
from ${{\bf v}}$ and $\dot{{\bf v}}$ alone, as it will contain $\ddot{{\bf v}}$-dependent terms too.
Thus by comparing the change in the mechanical power between the two cases (i.e., uniform acceleration and non-uniform 
acceleration cases), one should be able to calculate the excess power going in the fields above the actual rate of change of 
the kinetic energy of the charge. 

Laws of mechanics determine the actual charge motion at $t_{\rm o}$, taking $\ddot{{\bf v}}$ also into consideration 
\begin{eqnarray}
\label{eq:41e}
{\bf v}_{\rm o}={\bf v}+ \dot{{\bf v}} (t_{\rm o}-t')+ \frac{\ddot{{\bf v}}(t_{\rm o}-t')^2}{2}\:,
\end{eqnarray}
\begin{eqnarray}
\label{eq:41f}
\dot{{\bf v}}_{\rm o}=\dot{{\bf v}}+\ddot{{\bf v}} (t_{\rm o}-t').
\end{eqnarray}
The electrodynamics fields (Eq. (\ref{eq:1})) do not take into consideration any rate of change of acceleration. 
For a finite rate of change of acceleration, the velocity and thereby kinetic energy of charge at $t_{\rm o}$
would contain $\ddot{{\bf v}}$, meaning charge would have different kinetic energy than
what went into its electromagnetic fields, the latter not taking $\ddot{{\bf v}}$ into account. \\
The expression for the kinetic power is, 
\begin{equation}
\label{eq:41g0}
{\cal P}={\rm d}(m_{el} {\bf v}_{\rm o}^2 /2 )/{\rm d}t = m_{el} \dot{{\bf v}}_{\rm o} \cdot {\bf v}_{\rm o}\:, 
\end{equation}
which for a uniform acceleration case ($\ddot{{\bf v}}_{\rm o}=0$) 
from Eqs. (\ref{eq:41b}) and (\ref{eq:41c}) is, 
\begin{equation}
\label{eq:41g1}
{\cal P}_1=m_{el}\dot{{\bf v}}\cdot\bigg[{\bf v}+ \dot{{\bf v}} (t_{\rm o}-t')\bigg]. 
\end{equation}
The expression for the power going into the kinetic energy of the charge in a non-uniform acceleration case 
($\ddot{{\bf v}}_{\rm o} \ne 0$), from Eqs.~(\ref{eq:41e}), (\ref{eq:41f}) and (\ref{eq:41g0}) is
\begin{eqnarray}
\nonumber
{\cal P}_2&=&m_{el}\bigg[\dot{{\bf v}}+\ddot{{\bf v}}(t_{\rm o}-t')\bigg]\\
\label{eq:41g2}
&&\!\!\!\!\!\!\!\!\!\!\!\!\cdot\bigg[{\bf v}+ \dot{{\bf v}} (t_{\rm o}-t') +  \frac{\ddot{{\bf v}}(t_{\rm o}-t')^2}{2}\bigg].
\end{eqnarray}
But this is not the power going into the changing electromagnetic fields of the charge, which does not involve 
$\ddot{{\bf v}}$ (see Eq. (\ref{eq:1})) and is thus still given by  Eq.~(\ref{eq:1f3}), and equals ${\cal P}_1$ (Eq.~(\ref{eq:41g1})). 
The excess power, $\Delta {\cal P}$, going into the fields over and above the actual kinetic power of the charge (${\cal P}_1-{\cal P}_2$) then is,
\begin{eqnarray}
\nonumber
\Delta {\cal P}&=\;\;\;m_{el}\dot{{\bf v}}\cdot\bigg[{\bf v}+ \dot{{\bf v}} (t_{\rm o}-t')\bigg]\\
\nonumber
&-m_{el}\bigg[\dot{{\bf v}}+\ddot{{\bf v}}(t_{\rm o}-t')\bigg]\\
\label{eq:41g}
&\cdot \bigg[{\bf v}+\dot{{\bf v}} (t_{\rm o}-t') +  \frac{\ddot{{\bf v}}(t_{\rm o}-t')^2}{2}\bigg], 
\end{eqnarray}
which to the lowest order in $t_{\rm o}-t'$ ($=r_{\rm o}/c)$ is 
\begin{eqnarray}
\label{eq:41h}
\Delta {\cal P}=-m_{el} \ddot{{\bf v}} \cdot{\bf v}_{\rm o} (r_{\rm o}/c).
\end{eqnarray}
Substituting for the electric mass of a charge, $m_{el}=2e^2/3r_{\rm o} c^2$, the excess power in the electromagnetic fields is, 
\begin{eqnarray}
\label{eq:41i}
\Delta {\cal P}= -\frac{2e^2}{3r_{\rm o} c^2} \ddot{{\bf v}}\cdot{\bf v}_{\rm o}\frac{r_{\rm o}}{c}= \frac{- 2e^2 \ddot{{\bf v}}\cdot{\bf v}_{\rm o}}{3c^3}.
\end{eqnarray}
This is the formula for power losses from a radiating charge. 

We can write this power loss being due to a radiative drag force ${\bf F}$ as 
the charge moves with a velocity ${\bf v}_{\rm o}$.
\begin{eqnarray}
\label{eq:41j}
\Delta {\cal P}= -{\bf F}\cdot{\bf v}_{\rm o}= \frac{- 2e^2 \ddot{{\bf v}}\cdot{\bf v}_{\rm o}}{3c^3}.
\end{eqnarray}
or  
\begin{eqnarray}
\label{eq:41k}
\left[{\bf F}- \frac{2e^2 \ddot{\bf v}}{3c^3}\right]\cdot{\bf v}_{\rm o}=0.
\end{eqnarray}
Since in Eq. (\ref{eq:41k}) ${\bf v}_{\rm o}$ is an arbitrary vector, implying that the equation is true for all values 
of ${\bf v}_{\rm o}$, we have 
\begin{eqnarray}
\label{eq:41l}
{\bf F}= \frac{2e^2 \ddot{\bf v}}{3c^3}.
\end{eqnarray}
Here it could be objected that one could add to ${\bf F}$ any arbitrary vector ${\bf A}$ such that ${\bf v}_{\rm o} \times {\bf A}=0$, and 
still satisfy Eq. (\ref{eq:41k}). For instance, the force on a charge in a magnetic field ${\bf B}$ is proportional to ${\bf v}_{\rm o} \times {\bf B}$. 
However, since no power loss results in such a case, it does not represent any radiation reaction force. Therefore Eq. (\ref{eq:41l}) remains valid 
for the radiation reaction force. 

This formula for radiative drag force or radiation reaction is the same as derived in literature from the self-force of a charged sphere. 
But we have here derived radiation reaction and the radiative losses from the kinetic power of the electric inertial mass of a charged particle. 

It has sometimes been stated in the literature \cite{51} that radiation reaction is not represented 
correctly by Eq.~(\ref{eq:41l}), and that it should instead be calculated from Larmor's formula for radiative losses. However, that is an 
erroneous statement and by using synchrotron radiative losses as an example, it can be conclusively demonstrated that the radiation damping calculated from Larmor's formula (or its relativistic generalization 
Li\'{e}nard's formula) does not yield results compatible with the special relativity and further, violates energy-momentum conservation \cite{p58a}. 

In an assumedly uniform and homogeneous magnetic field, a charge will be moving in a helical path with a velocity component 
${\bf v}_{\parallel}={\bf v} \cos \psi$, parallel to the magnetic field, where $\psi$ is the pitch angle (i.e., angle with 
respect to the magnetic field vector) of the charge. Since the radiation is confined to a narrow cone 
around the instantaneous direction of motion of the charge \cite{1,25,27}, from Larmor's formula (or rather from Li\'{e}nard's formula), 
any radiation reaction on the charge will be in a direction just opposite to its instantaneous velocity vector \cite{23, 26}, 
implying no change in the pitch angle of the charge. Thus the ratio ${\bf v}_\perp/{\bf v}_\parallel = \tan\psi$, will not change. 
However, there is something amiss in the above arguments and the above picture is not consistent with the 
special theory of relativity. A more careful consideration shows that in the case of synchrotron losses, the velocity component parallel to 
the magnetic field (${\bf v}_\parallel$) of the charge remains unaffected, while magnitude of ${\bf v}_\perp$ steadily decreases due to 
radiative losses and as a consequence the pitch angle of the radiating charge in general changes, with the charge motion gradually 
getting aligned with the magnetic field 
direction \cite{p30}. Thus the dynamics of the charged particle computed from Larmor's formula (or its relativistic generalization 
Li\'{e}nard's formula) does not yield results compatible with the special relativity and that only the radiation reaction formula  
yields a picture consistent with the special relativity.

\section{Conclusions}
From  mechanical considerations of electric inertial mass of a charge, formulas for radiation reaction and radiative losses were derived, albeit in a heuristic manner. 
The derivation made use of the fact that,

1. A moving charge has an electromagnetic field momentum which we can infer from classical mechanics if one uses its electric inertial mass.

2. In the case of a uniformly accelerated charge, its rate of change in kinetic energy is concurrent with the rate of 
change in its electromagnetic field energy, and is given by the scalar product of its instantaneous velocity with the 
rate of change of its electromagnetic momentum.

3. In the case of a varying acceleration, the energy in the electromagnetic fields changes at a different rate than that of 
change of kinetic energy of the charge and it is this energy difference that is not represented in the actual motion of the charge 
and can be called as a radiative loss.

The accordingly derived instantaneous power loss turns out to be directly proportional 
to the scalar product of the velocity and the rate of change of acceleration of the charge as derived earlier in literature from radiation 
reaction due to the self-force of the charge. 
{}
\end{document}